\documentclass[peerreview,draftcls,doublespace,oneside,onecolumn,11pt]{IEEEtran}

\usepackage{graphicx,cite,amssymb,epsfig,amsmath}
\usepackage[large]{caption}

\begin{document}
%
\title{Survey of Cognitive Radio Techniques in Wireless Network}

\author{Lu Lu
           }
\maketitle

\thispagestyle{plain}
\setcounter{page}{1}

Wireless access networks were originally developed for different communication scenarios. Wireless networks aim to provide ubiquitous, flexible
communications mainly in community areas~\cite{5,8,10,11}. In wireless network, there always exists the need for the communication services at higher data rates and quality of service (QoS)~\cite{1,2,3,4}. Thus lots of different techniques have been explored to improve modern wireless networks~\cite{12,13} to satisfy different requirements and combat different problems. For example, the rapid growth in the ubiquitous wireless services has imposed increasing stress on the fixed and limited radio spectrum. Thus allocating a fixed frequency band to each wireless service, which is the current frequency allocation policies, is a necessary to eliminate interference between different wireless services~\cite{6,8,10}. As a result, innovative techniques that can offer new ways of exploiting the available spectrum are needed~\cite{1,17,18}. Hence, cognitive radio~\cite{1,7} arises as a feasible solution to the aforementioned spectral resource allocation problem by introducing the opportunistic usage of the frequency bands that are not heavily occupied by licensed users. Currently there are many different methods for spectrum sensing have been proposed, such as the matched filtering approach~\cite{15}, the feature detection approach~\cite{16} and the energy detection approach~\cite{17}. For the matched filtering method, it can maximize the SNR inherently. However it is difficult to do detection without signal information such as pilot and frame structure. And for feature detection method which is basically performed based on cyclostationarity, it also must have information about received signal sufficiently. However, in practice, cognitive radio system can not know about primary signal¡¯s structure and information. For the energy detection method, although it doesn't need any information about the signal to be detected, it is prone to false detections since it is only based on the signal power~\cite{18}. When the signal is heavily fluctuated or noise uncertainty is big, it becomes difficult to discriminate between the absence and the presence of the signal. In addition, the energy detection is not optimal for detecting the correlated (colored) signals, which are often found in practice. Thus we can see that different methods have different advantages and disadvantages, one can choose to use these different methods according to the different requirements for specific problems.

\bibliographystyle{IEEEtran}


\end{document}